\documentclass[12pt]{article}
\usepackage{graphicx}

      \usepackage[cp1251]{inputenc} 

\begin{document}


 \vskip 1.0cm
 \begin{center} {\LARGE\bf Galactic Orbits of Hipparcos Stars: Classification of Stars}
  \end{center}

 \bigskip
 \centerline {\bf G. A. Gontcharov and A. T. Bajkova}
 \medskip
\centerline{\small\it
 Pulkovo Astronomical Observatory, Russian Academy of Sciences,
 St.-Petersburg, Russia
 }

 \bigskip
{\bf Abstract}---The Galactic orbits of 27 440 stars of all
classes with accurate coordinates and parallaxes of more than 3
mas from the Hipparcos catalogue, proper motions from the Tycho-2
catalogue, and radial velocities from the Pulkovo Compilation of
Radial Velocities (PCRV) are analyzed. The sample obtained is much
more representative than the Geneva–Copenhagen survey and other
studies of Galactic orbits in the solar neighborhood. An
estimation of the influence of systematic errors in the velocities
on orbital parameters shows that the errors of the proper motions
due to the duplicity of stars are tangible only in the statistics
of orbital parameters for very small samples, while the errors of
the radial velocities are noticeable in the statistics of orbital
parameters for halo stars. Therefore, previous studies of halo
orbits may be erroneous. The distribution of stars in
selection-free regions of the multidimensional space of orbital
parameters, dereddened colors, and absolute magnitudes is
considered. Owing to the large number of stars and the high
accuracy of PCRV radial velocities, nonuniformities of this
distribution (apart from the well-known dynamical streams) have
been found. Stars with their peri- and apogalacticons in the disk,
perigalacticons in the bulge and apogalacticons in the disk,
perigalacticons in the bulge and apogalacticons in the halo, and
perigalacticons in the disk and apogalacticons in the halo have
been identified. Thus, the bulge and the halo are inhomogeneous
structures, each consisting of at least two populations. The
radius of the bulge has been determined: 2 kpc.

\section{INTRODUCTION}

Using the stellar coordinates $\alpha$ and $\delta$, parallaxes
$\pi$ from the Hipparcos catalogue (van Leeuwen 2007), proper
motions $\mu$ from the Tycho-2 catalogue (H\"{o}g et al. 2000),
and radial velocities $V_{r}$ allows not only the complete set of
coordinates $X$, $Y$, $Z$ and velocity components $U$, $V$, $W$,
but also the Galactic orbits of stars to be calculated.

Because of the small number of stars with accurate $V_{r}$ and
because of the distrust in the joint use of $\mu$ and $V_{r}$, the
Galactic orbits have been investigated so far only for small lists
of stars, as a rule, obtained with the same instrument. Since the
samples are incomplete, the studies of the orbits for stars that
do not belong to the Galactic disk are particularly poor.

The appearance of the Pulkovo Compilation of Radial Velocities for
35 493 Hipparcos stars (PCRV; Gontcharov 2006), in which the
systematic errors of $V_{r}$ were taken into account and all the
main classes of stars are represented, allows one to set the task
of comprehensively studying the statistical characteristics of
samples of stars based on their Galactic orbits by invoking their
metallicities and ages.

The PCRV is still the largest source of $V_{r}$ with included
systematic errors. The median accuracy of $V_{r}$ from the PCRV is
0.7 km s.1; $V_{r}$ is more accurate than 5 km s.1 for all stars.
The PCRV includes the values of $V_{r}$ from 203 catalogues for
which the systematic errors detected in them were taken into
account. These include the two largest present-day catalogues: the
Geneva–Copenhagen survey (GCS) of more than 14 000 stars mostly of
types FV–GV near the Sun (Nordstr . om et al. 2004; Holmberg et
al. 2007, 2009) and the kinematic survey of more than 6000
KIII–MIII stars based on CORAVEL observations (Famaey et al.
2005). The velocity components and Galactic orbits for 11 218
stars calculated in the GCS are used in this study to check the
results, while the metallicities Fe/H for 11 615 stars and the
ages for 10 249 stars calculated in the GCS will be used in our
subsequent studies of the age–kinematics and
metallicity–kinematics relations for various groups of stars.

\section{DATA REDUCTION}

The sample considered below is limited in parallax, $\pi>3$
milliarcseconds (mas), for the following reasons. No firm
conclusions can be reached about regions far from the Sun, where
the PCRV includes few stars, while the limitation $\pi<3$, leaving
most of the stars in the sample (28 600, 80.6\%), corresponds to a
space where the PCRV is sufficiently representative. In addition,
the accuracy of the components $\mu_{\alpha}\cos\delta$ and
$\mu_{\delta}$ for the overwhelming majority of PCRV stars is
higher than 3 mas yr$^{-1}$. This value corresponds to the
accuracy of $V_{r}$ used (higher than 5 km s$^{-1}$) at a distance
of 333 pc. The limitation of the sample at this distance makes it
homogeneous with regard to the accuracy of the velocity components
$U$, $V$, $W$. It is also important that, according to our
numerical simulations (see Gontcharov 2012a), the Lutz–Kelker and
Malmquist biases are negligible under the mentioned limitation in
$\pi$. These are the biases of the sample’s statistical
characteristics, primarily the distances and absolute magnitudes
of the stars being determined, that arise when the sample is
limited in measured parallax and/or in observed magnitude
(Perryman 2009, pp. 208–212).

In addition to $\pi>3$ mas, we adopted the following limitations:
the relative accuracy is $\sigma(\pi)/\pi<0.5$ (59 stars were
lost), the accuracies of the components are
$\sigma(\mu_{\alpha}\cos\delta)<5$ and $\sigma(\mu_{\delta})<5$
mas yr$^{-1}$ (37 stars were lost), and the accuracies of the
Tycho- 2 photometry are $\sigma(B_{T})<0.1^{m}$ and
$\sigma(V_{T})<0.1^{m}$(243 stars were lost). The final sample
contains 27 440 stars.

More severe limitations, for example, $\sigma(\pi)/\pi<0.2$
alongwith $\sigma(\mu_{\alpha}\cos\delta)<3$ and
$\sigma(\mu_{\delta})<3$ mas yr$^{-1}$, leave 25 082 stars in the
sample. In this case, all of the conclusions reached in this study
remain valid and all of the categories of stars found are
identified with no lesser confidence. However, a considerable
number of bulge and halo stars that are few anyway are lost under
severe limitations. Therefore, here we give preference to the
mentioned mild limitations to exclude only the stars whose data do
not allow them to be classified.

The dereddened color $(B_{T}-V_{T})_{0}$ was calculated for each
star:
\begin{equation}
\label{btvt0} (B_{T}-V_{T})_{0}=(B_{T}-V_{T})-E(B_{T}-V_{T}),
\end{equation}
where the reddening
$E(B_{T}-V_{T})=A_{V_{T}}/R_{VT}\approx1.1A_{V}/1.2R_{V}$. The
coefficients in this formula were calculated by taking into
account the extinction law from Draine (2003); the extinction
$A_{V}$ was calculated from our 3D analytical extinction model
(Gontcharov 2009, 2012b) as a function of the trigonometric
distance $r=1/\pi$. and Galactic coordinates $l$ and $b$, while
the extinction coefficient $R_{V}$ was calculated from the 3Dmap
of its variations as a function of the same coordinates
(Gontcharov 2012a).The absolute magnitude $M_{V_{T}}$ was
calculated for each star from the formula
\begin{equation}
\label{mvt} M_{V_{T}}=V_{T}+5-5\lg~r-A_{V_{T}}.
\end{equation}

The positions of our 27 440 sample stars on a Hertzsprung–Russell
(H–R) diagram of the form ``$(B_{T}-V_{T})_{0}$ -- $M_{V_{T}}$''
are shown in Fig. 1a. The sample under consideration contains
almost all GCS stars with accurate data and a similar diagram for
them is shown in Fig. 1b. It can be seen that, following the PCRV
and in contrast to the GCS, all classes, including the main
sequence (MS), the red giant clump and branch, supergiants,
subgiants, subdwarfs, red dwarfs, and one white dwarf, are
represented in the sample. The line indicates the theoretical
zero-age main sequence (ZAMS) from the Padova database of
evolutionary tracks and isochrones (http://stev.oapd.inaf.it/cmd;
Bressan et al. 2012), which was fitted with an accuracy (about
0.1m) sufficient for the subsequent analysis by the polynomial
\begin{equation}
\label{zams} Y=5.9X^5-19.34X^4+21.1X^3-8.8X^2+5.8X+1.7,
\end{equation}
where $X=(B_{T}-V_{T})_{0}$, $Y=M_{V_{T}}$. The cross indicates
typical errors, $\sigma((B_{T}-V_{T})_{0})=0.02^m$ and
$\sigma(M_{V_{T}})=0.5^m$, for an individual star. It can be seen
that cloud of points of MS stars is located mainly to the right
and above the ZAMS, as it must be. However, in the range
$0^m<(B_{T}-V_{T})_{0}<0.35^m$, which roughly corresponds to the
spectral type AV, the well-known deviation of the cloud from the
shown ZAMS is noticeable. This deviation will be discussed in a
separate study.

The Galactic orbits of the stars under consideration were
calculated using the Galactic potential from Fellhauer et al.
(2006) and Helmi et al. (2006):
$\Phi=\Phi_{halo}+\Phi_{disk}+\Phi_{bulge}$. In this case,

• the halo was represented by a potential dependent on the
cylindrical Galactic coordinates $R$ and $Z$ as
$\Phi_{halo}(R,Z)=\nu_0^2\ln(1+R^2/d^2+Z^2/d^2)$, where
$\nu_0=134$ km s$^{-1}$ and $d=12$ kpc;

• the disk was represented by the potential from Miyamoto and
Nagai (1975) as a function of the same coordinates:
$\Phi_{disk}(R,Z)=-GM_d(R^2+(b+(Z^2+c)^{1/2})^2)^{-1/2}$, where
the disk mass $M_{d}=9.3\cdot10^{10}$ $M_{\odot}$, $b=6.5$ kpc,
and $c=0.26$ kpc;

• the bulge was represented by the potential from Hernquist
(1990): $\Phi_{bulge}(R)=-GM_{b}/(R+a)$, where the bulge
$M_{b}=3.4\cdot10^{10}$ $M_{\odot}$ and $a=0.7$ kpc.

The Galactic rotation velocity at $r=8$ kpc for the Sun was taken
to be 220 km s$^{-1}$. The solar motion relative to the local
standard of rest was taken to be ($U=10$, $V=11$, $W=7$) km
s$^{-1}$ based on the results by Bobylev and Bajkova (2010) in
agreement with the results by Sch$\ddot{o}$nrich et al. (2010) and
Gontcharov (2012d).

The key characteristics of the calculated orbits are the peri- and
apogalactic distances designated below as  $r_{min}$ and
$r_{max}$, respectively, the orbital eccentricity $e$, and the
largest distance of the orbit from the Galactic plane $Z_{max}$.

\section{ESTIMATING THE INFLUENCE OF ERRORS}

Let us estimate the influence of errors in $V_{r}$ and $\mu$ on
$r_{min}$, $r_{max}$, $e$ and $Z_{max}$.

For double and multiple stars, the observed component or
photocenter can move over the celestial sphere nonlinearly. For
visual binary stars (resolved systems), this occurs due to the
orbital motion of the components or because one of them falls or
does not fall within the field of view. For astrometric binaries
(unresolved systems), this occurs due to the orbital motion of the
system’s photocenter relative to the barycenter. In both cases,
the variability of at least one of the components can also have an
effect. Although Vityazev et al. (2003) showed an insignificant
influence of the orbital motions in star pairs on the stellar
kinematics, let us estimate this influence by a different method.

The proper motions from catalogues with a large difference of
epochs (e.g., more than 50 years for Tycho-2), a small difference
of epochs (e.g., 3.5 years for Hipparcos), and the orbital motions
were compared, for example, by Gontcharov et al. (2001) and
Gontcharov and Kiyaeva (2002). It follows from this comparison
that precisely the orbital motions are responsible for the large
differences of $\mu$ in such catalogues as Hipparcos and Tycho-2
and that here it is more appropriate to use $\mu$ from Tycho-2.
Their replacement by . from Hipparcos when calculating the
Galactic orbits just reflects the influence of the orbital motions
in star pairs. The standard deviation of the differences between .
from Tycho-2 and Hipparcos for the stars under consideration is
1.8 mas yr$^{-1}$. The values of  $r_{min}$, $r_{max}$, $e$ and
$Z_{max}$  calculated using $\mu$ from Tycho-2 and $r'_{min}$,
$r'_{max}$, $e'$ and $Z'_{max}$ calculated using $\mu$  from
Hipparcos differ insignificantly: the standard deviations of the
differences are $\sigma(r'_{max}-r_{max})=0.15$ kpc,
$\sigma(r'_{min}-r_{min})=0.02$, $\sigma(e'-e)=0.01$,
$\sigma(Z_{max}-Z'_{max})=0.02$ kpc. Thus, although the Galactic
orbital parameters for an individual star can be strongly affected
by inaccurate/incomplete knowledge of its duplicity, the latter
does not affect the statistical results of analyzing the Galactic
orbits for a sample of ten or more stars.

When creating the PCRV, we found and took into account numerous
and various physically justified dependences of $V_{r}$ on it
itself (probably primarily due to the scale error and the tilt of
the focal plane when the distances between spectral lines were
measured), on the $(B-V)$ color (probably primarily due to the
differences between the measured and comparison spectra), and on
the celestial coordinates $\alpha$ and $\delta$ (probably
primarily due to the seasonal correlations between the celestial
coordinates, the instrument’s temperature, and the hour angle) in
the original catalogues. For the original catalogues produced with
similar instruments, similar systematical dependences were found
in the PCRV. For example, the GCS and the results by de Medeiros
and Mayor (1999) obtained with the identical CORAVEL spectrometers
give coincident (within the accuracy limits) dependences of
$V_{r}$ on $V_{r}$, $(B-V)$, and $\alpha$, as shown in Fig. 1 and
Table 3 from Gontcharov (2006).

To model the influence of systematic errors in $V_{r}$, we
calculated the orbits using a radial velocity distorted by errors:
\begin{equation}
\label{gcs}
V'_{r}=V_{r}+2.52(B-V)^2-4.7(B-V)+0.002V_{r}-0.034\alpha-0.04\cos(\alpha+0.7)+1.88,
\end{equation}
where $V'_r$ and $V_{r}$ are in km s$^{-1}$, $(B-V)$ is in
magnitudes, and $\alpha$ is in radians. This formula reflects the
errors found by Gontcharov (2006) in the GCS (an error was made in
the text of the paper when the PCRV was published but not in the
calculations: there should be $\cos(\alpha+0.7)$ or
$\cos(\alpha+40^{\circ})$ if the phase is in degrees instead of
the published $\cos(\alpha-40^{\circ})$).

For the stars under consideration, the difference $|V'_{r}-V_{r}|$
from Eq. (4) can reach 2.3 km s$^{-1}$, which is approximately
triple the median random error of $V_{r}$ in the PCRV.

In our analysis, it makes sense to model the errors by Eq. (4)
instead of using $r_{min}$, $r_{max}$, $e$ and $Z_{max}$ directly
from the GCS, because these errors, as follows from our comparison
of the GCS and the results by de Medeiros and Mayor (1999), also
extend to the spectral types of stars earlier than F and later
than G, which are virtually absent in the GCS. In addition, the
GCS results differ from the results of this study due to the
differences not only in radial velocities but also in distances
and extinction estimates, while we want to estimate the influence
of only the errors in $V_{r}$.

The values of $r_{min}$, $r_{max}$, $e$ and $Z_{max}$ calculated
using $V_{r}$ and $r'_{min}$, $r'_{max}$, $e'$ and $Z'_{max}$r
calculated using $V'_{r}$ differ insignificantly for disk stars
and markedly for halo stars, i.e., for stars with $r_{max}>20$ and
$Z_{max}>4$. Consequently, the systematic errors of $V_{r}$ affect
the statistical results of analyzing the Galactic orbits of only
the halo stars due to the small number of these stars in the
sample. Thus, either a careful allowance for the systematic errors
in $V_{r}$ or sample completeness is needed to analyze the
Galactic orbits of halo stars. Obviously, both conditions are
violated in most of the previous studies of the halo orbits.

The mean value of the differences $\overline{r'_{max}-r_{max}}$,
just as of the remaining orbital parameters, is zero. The standard
deviations of the differences are $\sigma(r'_{max}-r_{max})=0.44$
kpc, $\sigma(r'_{min}-r_{min})=0.03$ kpc, $\sigma(e'-e)=0.01$,
$\sigma(Z'_{max}-Z_{max})=0.08$ kpc. A direct comparison of the
orbital parameters from the GCS with those obtained here gives the
mean values of the differences $\Delta r_{max}=0.19$ kpc, $\Delta
r_{min}=0.08$ kpc, $\Delta e=0.005$, $\Delta Z_{max}=-0.05$ kpc
and their standard deviations $\sigma(\Delta r_{max})=0.35$ kpc,
$\sigma(\Delta r_{min})=0.19$ kpc, $\sigma(\Delta e)=0.02$,
$\sigma(\Delta Z_{max})=0.15$ kpc. The noticeable differences
between the modeling and the real GCS for
$\sigma(r'_{min}-r_{min})=0.03$ versus $\sigma(\Delta
r_{min})=0.19$ kpc and $\sigma(Z'_{max}-Z_{max})=0.08$ versus
$\sigma(\Delta Z_{max})=0.15$ kpc are caused by the differences in
distance and extinction estimates in this study compared to the
GCS.

Here, we considered the main sources of systematic errors and, in
addition, can estimate the random errors of the orbital parameters
based on the errors in the observed quantities. In our subsequent
analysis of the results, typical total error estimates are marked
in the figures. They are determined mainly not by the random
errors of the velocities but by their systematic errors, which
should not be ignored.

\section{RESULTS}

Figures 1c and 1d show the positions of the sample and GCS stars,
respectively, on the ``$(B_{T}-V_{T})_{0}$ -- $e$'' diagram; Figs.
1e and 1f, 1g and 1h, and 1i and 1j show their positions on the
``$(B_{T}-V_{T})_{0}$ -- $r_{max}$'', ``$(B_{T}-V_{T})_{0}$ --
$r_{min}$'', and ``$(B_{T}-V_{T})_{0}$ -- $Z_{max}$'' diagrams,
respectively (the orbital parameters in Figs. 1d, 1f, 1h, and 1j
were taken directly from the GCS).

It can be seen that all categories of stars except FV–GV are
represented in the GCS much more poorly than they are in the
sample under consideration, although the distribution has the same
structure.

The dashed lines in Figs. 1c, 1e, 1g, and 1i indicate an
approximate separation of the stars into six categories by their
evolutionary status and membership in Galactic subsystems. Below,
we analyze separately

• 5377 stars with $(B_{T}-V_{T})_{0}<0.3^m$, below referred to as
the OA subsample),

• 12 600 stars with $0.3^m<(B_{T}-V_{T})_{0}<0.85^m$ (mostly
FV–GV, below referred to as the FG subsample),

• 9463 stars with $(B_{T}-V_{T})_{0}>0.85^m$ (mostly K–M with a
large fraction of giants, below referred to as the KM subsample).

The results for the FG subsample can be compared with those from
the GCS. Each subsample was divided in the figure by the
horizontal dashed line approximately into stars with eccentric
orbits (halo and bulge stars) and with more circular orbits (disk
stars, so far without any separation into the thin and thick
disks, given that this separation has recently been called into
question (Bovy et al. 2012)).

The present-day theory of stellar evolution suggests that hot
subdwarfs (evolved stars mainly with reactions in the helium core,
i.e., on the horizontal giant branch), cool subdwarfs (unevolved
low metallicity dwarfs near the MS) and low-mass branch giants
dominate among the high-eccentricity stars of the OA, FG, and KM
subsamples, respectively. It can be seen that there are much more
high-eccentricity stars in the FG subsample than in the remaining
subsamples. This could not be affected by the sample selection in
$r$ and $V_{T}$, because the absolute magnitudes of hot and cool
subdwarfs are approximately equal, while giants are seen even at a
greater distance. Most of the high-eccentricity stars we observe
theoretically have a mass smaller than 0.8 Solar mass. Therefore,
the predominance of cool subdwarfs is explained by the fact that
only the oldest and low-mass metal-poor stars have managed to
become giants in the lifetime of the Galaxy, while the majority
remain near the MS.

For nonsingle stars in Hipparcos, the orbital parameters, $\pi$
and $\mu$ were determined jointly from directly observed
quantities, the abscissas on the reference great circle. The
orbital motion can distort the observed $\pi$ predominantly in the
direction of its increase. When the angular measure is converted
to the linear one, the measured $\mu$ then turn out to be smaller
than the true ones due to the erroneous distance. As a result, the
calculated e is larger than the true one and disk stars
demonstrate the orbits of halo or bulge stars. This effect should
manifest itself irrespective of $(B_{T}-V_{T})_{0}$. However, the
significantly nonuniform distribution of halo and bulge stars on
$(B_{T}-V_{T})_{0}$ seen in Fig. 1 suggests that this effect is
absent.

The following effects are less pronounced in Fig. 1 but important.
The small number of high-eccentricity stars in the range
$0.1^m<(B_{T}-V_{T})_{0}<0.25^m$ corresponds to the theory of
subdwarfs (Gontcharov et al. 2011). A local minimum in the scatter
of orbital parameters for disk stars is reached at
$(B_{T}-V_{T})_{0}\approx0.1^m$, while this scatter in the range
$0.3^m<(B_{T}-V_{T})_{0}<0.4^m$ increases sharply.

Applying the more stringent star selection criteria
$\sigma(\pi)/\pi<0.2$, $\sigma(\mu_{\alpha}\cos\delta)<3$ and
$\sigma(\mu_{\delta})<3$  mas yr$^{-1}$, which leave only 25 082
stars in the sample, does not change qualitatively the
distribution of stars on the diagrams of Figs. 1a, 1c, 1e, 1g, and
1i. As an example, Fig. 2 presents the ``$(B_{T}-V_{T})_{0}$ --
$r_{min}$'' (a) and ``$(B_{T}-V_{T})_{0}$ -- $Z_{max}$'' (b)
diagrams for 25 082 sample stars with the above stringent
selection criteria, which are worth comparing with Figs. 1g and
1i, respectively.

\medskip

{\it Comparison of Orbital Parameters}

\medskip

Figure 3 shows the distribution of stars from the OA subsample on
the (a) ``$r_{max}$ -- $e$'', (b) ``$Z_{max}$ -- $e$'', (c)
``$r_{min}$ -- $e$'',  (d) ``$r_{min}$ -- $Z_{max}$'', (e)
``$r_{max}$ -- $r_{min}$'', and (f) ``$r_{max}$ -- $Z_{max}$''
diagrams. Here, the limitations due to the sample selection by $r$
and $V_{T}$ are noticeable. For example, the selection in Fig. 3c
manifests itself in the correlation between $r_{min}$ and $e$.
Despite the selection, some conclusions are possible.

As expected, the overwhelming majority of stars from the OA
subsample are young low-eccentricity stars. Indeed, Gontcharov
(2012d) found a correlation between the age and
$(B_{T}-V_{T})_{0}$ for stars near the MS with
$(B_{T}-V_{T})_{0}<0.7^m$:
\begin{equation}
\label{exp2} T=0.42e^{3.86(B_{T}-V_{T})_{0}},
\end{equation}
which agrees well with that from the GCS data. According to this
formula, the OA-subsample stars near the MS ($>80\%$ of the
subsample) have ages younger than 1.34 Gyr and nearly circular
orbits ($e<0.1$, $r_{max}<10$ kpc, $r_{min}>6$ kpc, $Z_{max}<0.3$
kpc) and, thus, belong to the thin disk. A nonuniformity of the
distribution of these stars is noticeable in Fig. 3e. It is caused
primarily by selection in favor of nearby stars and, accordingly,
by a predominance of stars with $r_{min}\approx8$ and
$r_{max}\approx8$ kpc. The additional star density variations in
the densely populated region of Fig. 3e are caused by the
influence of the dynamical streams of disk stars or super clusters
considered, for example, in the GCS, by Famaey et al. (2005) and
Gontcharov (2012c, 2012d). The orbits of disk stars will be
analyzed in detail separately.

The stars with $r_{min}<2$ kpc and $e>0.6$ are isolated in Fig. 3c
in these parameters from the remaining ones and probably belong to
the bulge. They are clearly divided into two groups:
$7.8<r_{max}<9.5$ kpc and $10.5<r_{max}<12.2$ kpc. Selection could
not eliminate the stars with intermediate $r_{max}$. Consequently,
there is a physical cause of their absence. The separation of
bulge stars into at least two groups should be admitted. However,
there are different $Z_{max}$ and $Fe/H$ in each of these groups
and the stars from the groups occupy approximately the same region
of the H–R diagram, being known or suspected hot subdwarfs (the
blue part of the horizontal branch). The only detected differences
are related to the difference in $r_{max}$: the group with larger
$r_{max}$ has larger absolute values of the velocity component
$|U|>150$ km s$^{-1}$ and is in the longitude octants of the
Galactic center and anticenter, while the group with smaller
$r_{max}$ has $|U|<150$ km s$^{-1}$ and is at longitudes far from
the center–anticenter line. The distribution of bulge stars in
apogalactic distances are possibly subjected to density waves and
other dynamical processes, as are the dynamical, predominantly
radial Sirius (Ursa Major), Pleiades, Hyades, Coma Berenices,
$\alpha$ Cet/Wolf 630, Hercules, and other steams (Gontcharov
2012c, 2012d). We then see two groups of bulge stars that are
dynamically associated with two spiral arms. Judging by the mean
$r_{max}$ for the stars from these groups that are not too far
from the Galactic plane, these arms are at Galactocentric
distances of 8.3 (the arm near the Sun) and 11.6 kpc,
respectively. Obviously, the group of stars with $3.3<r_{min}<4.5$
kpc and $0.3<e<0.42$ is then also isolated in Fig. 3c not by
selection but by the dynamical processes that associated it with
the arm nearest to us ($\overline{r_{max}}=8.3$ kpc) and possibly
the arm or a different structure near the perigalacticons of these
stars.

Let us analyze the entire FG subsample and only the GCS stars from
this subsample, respectively, in Figs. 4 and 5, which are similar
to Fig. 3. The orbital parameters of the stars from the GCS rather
than those that we calculated are used in Fig. 5.

Just as in the OA subsample, most of the stars in the FG subsample
have  $e<0.3$, $6<r_{min}<8$ kpc, $8<r_{max}<10$ kpc, and
$Z_{max}<0.5$ kpc, i.e., these are disk stars with moderately
eccentric orbits. Formula (5), which gives ages from 1.34 to 11
Gyr for these stars, is valid for them.

As can be seen from our comparison of Figs. 4 and 5, the remaining
stars, i.e., those with comparatively eccentric orbits, are
represented in this study much better than in the GCS. A
nonuniform distribution of stars with $e>0.4$, $Z_{max}>3$ kpc,
and $r_{max}>15$ kpc on the graphs of Fig. 4 is particularly
noticeable. Because of the selection effect, not so much the
regions of enhanced star density in the figure as the voids
between them that are surrounded by stars and, therefore, that did
not result from selection are important.

Judging by the almost complete absence of stars with $2<Z_{max}<3$
kpc, the boundary between the disk and halo as, respectively, the
subsystem consisting mostly of stars born in the Galaxy itself and
the subsystem with a significant fraction of stars accreted from
the Galaxy’s disrupted satellites is clearly seen here. In support
of this, we see a comparatively uniform distribution of disk stars
($2<Z_{max}<3$ kpc) and a nonuniform distribution of halo stars on
the graphs. Two groups are identified with confidence among the
latter:
\begin{description}
\item[\normalfont{the bulge–halo group with }] $e>0.8$,
$r_{min}<1.5$ kpc, $3<Z_{max}<13$ kpc, \item[\normalfont{the halo
group with}] $0.25<e<0.75$, $7<r_{min}<8$ kpc, < rmin < 8 kpc,
$13<r_{max}<60$ kpc, $10<Z_{max}<21$ kpc,
\end{description}
and two more groups are identified with lesser confidence:
\begin{description}
\item[\normalfont{group 3 with}] $0.65<e<0.8$, $3.8<Z_{max}<5$
kpc, $1.5<r_{min}<3.6$ kpc, \item[\normalfont{group 4 with}]
$0.4<e<0.55$, $3<Z_{max}<4.2$ kpc, $2.5<r_{min}<4.3$ kpc.
\end{description}

The uncertainty in identifying the groups stems from the fact that
selection allows one to delineate not the boundaries of the groups
but rather the boundaries of the voids between them. The
bulge–halo group includes stars with their perigalacticons in the
bulge and apogalacticons in the halo; the halo group includes
typical halo stars with moderately eccentric orbits (at higher
$e$, the perigalacticon falls into the bulge). Selection probably
hid the halo stars with approximately circular orbits whose
existence is not ruled out. Judging by the absence of intermediate
stars with $1.5<r_{min}<7$ kpc, these groups are fairly isolated.
Consequently, the stars with their apogalacticons in the halo can
be divided into at least two groups. The difference in origin may
be responsible for this division: the bulge–halo group was born in
the Galaxy, while the halo group was accreted from disrupted
satellites. The halo group is discussed below.

In the subsample under consideration, the bulge stars ($r_{min}<2$
kpc) are much more numerous than in the GCS and are isolated from
the remaining ones by voids on the graphs (though selection
probably hid the bulge stars with less eccentric orbits). In
addition to the mentioned bulge–halo group, a bulge–disk group
with $Z_{max}<2$ 2 kpc, $r_{min}<2$ kpc, and $e>0.6$, i.e., with
the apogalacticons in the disk, is identified among the stars with
their perigalacticons in the bulge. Thus, we see the separation of
the bulge stars into stars with disk and halo kinematics. Of
particular importance is the absence of stars with intermediate
$2<Z_{max}<3$ kpc between these groups, which cannot be the result
of selection, i.e., the Galactic bulge combines the properties of
the classical and disk-shaped bulges observed in other galaxies.
It has an inhomogeneous structure and is probably a mixture of
stars of different subsystems, as is hypothesized by Bensby et al.
(2013) and in the papers cited by them.

Let us analyze the KM subsample in Fig. 6, which is similar to
Figs. 3 and 4. Just as for the previous subsamples, most of the
stars here ($r_{min}\approx8$ kpc and $r_{max}\approx8$ kpc) are
disk stars with nearly circular orbits ($e<0.3$, $Z_{max}<2$ kpc),
many of which belong to the dynamical streams or superclusters
considered by Famaey et al. (2005). These streams manifest
themselves particularly clearly in the nonuniform distribution of
stars in the most densely populated part of Figs. 6a and 6e.

Just as for the FG subsample, a clear separation of the disk and
halo at $Z_{max}\approx3$ kpc and the separation of bulge stars (
(($r_{min}<2$ kpc) into stars with disk ($Z_{max}<1.5$ kpc) and
halo ($Z_{max}>4$ kpc) kinematics when only one star
($Z_{max}=2.7$ kpc) is present in the interval can be seen for the
KM subsample.

\medskip

{\it Orbital Parameters and Luminosity}

\medskip

Some categories of stars can be revealed by the deviation of their
$M_{V_{T}}$ from the ZAMS specified by Eq. (3). We will consider
this deviation from the ZAMS .MVT to be negative for
high-luminosity stars (e.g., branch giants) and positive for low
luminosity stars (subdwarfs and white dwarfs).

Figure 7 shows the distribution of stars from the OA and KM
subsamples on the following diagrams: ``$\Delta M_{V_{T}}$ --
$e$'' (a) and (b), respectively; ``$\Delta M_{V_{T}}$ --
$r_{max}$'' (c) and (d), respectively; ``$\Delta M_{V_{T}}$ --
$r_{min}$'' (e) and (f), respectively; ``$\Delta M_{V_{T}}$ --
$Z_{max}$'' (g) and (h), respectively. The solid vertical straight
line marks .MVT = 0m and the dashed straight line marks .MVT = 1m,
which corresponds to a 2. error in calculating $M_{V_{T}}$.

The star HIP 14754 with .MVT = 8.1m is the only white dwarf in the
sample and is marked in the figure by the large circle.

The hot subdwarfs known from spectroscopy are marked in Figs. 3a,
3c, 3e, and 3g by the squares. Those of them that belong to the
bulge ($r_{min}<2$ kpc, $e>0.6$) and two stars in the adjacent
regions on the graphs ($2<r_{min}<3.5$ kpc, $0.4<e<0.6$) have a
high luminosity ($\Delta M_{V_{T}}<0^m$). As has been pointed out
above, although their perigalacticons are in the bulge, judging by
their kinematics they belong to the disk ($Z_{max}<2.5$ kpc) or
halo ($Z_{max}>5$ kpc). On the contrary, the remaining known hot
subdwarfs have a low luminosity ($\Delta M_{V_{T}}>1^m$) and
belong to the disk ($e<0.4$, $Z_{max}<1.5$ kpc). Following
Gontcharov et al. (2011), this study confirms the existence of hot
subdwarfs belonging to the Galactic thin disk in the solar
neighborhood.

The hot subdwarfs are mostly stars of the horizontal branch. Its
location on the H–R diagram depends on stellar metallicity, age,
and mass. At solar metallicity, an age younger than $10\times10^9$
yr, and a mass of more than $0.8 M_{\odot}$, the horizontal branch
appears as a clump of giants at $0.9^m<(B_{T}-V_{T})_{0}<1.5^m$.
The horizontal branch becomes bluer with decreasing metallicity,
increasing age, and decreasing mass to the extent that it turns
out to be near the MS (blue horizontal branch) or even
considerably bluer and lower than the MS(extremely blue horizontal
branch) at $Fe/H<-0.8$, an age older than $13\times10^9$ yr, and a
mass of less than  $0.7 M_{\odot}$. The corresponding stars are
classified as sdA, sdB, and sdO. Thus, the main path of a star’s
transformation into a hot subdwarf is a very low metallicity or an
age older than that of the Galactic disk or very fast evolution
due to the large mass loss on the giant branch. The less common
paths of fast evolution of low-mass stars into the region of hot
subdwarfs are related to mass transfer in a close star pair or
other ways of the addition or removal of stellar material (for
references, see Gontcharov et al. 2011). In any case, the
existence of hot subdwarfs in the Galactic thin disk is far from
an unambiguous explanation.

The group of thin-disk BV stars exhibiting a low luminosity that
stands out in the figure with $\Delta M_{V_{T}}>1^m$,
$e\approx0.05$, $r_{max}\approx8.9$ kpc, $r_{min}\approx7.8$ kpc,
$Z_{max}\approx0.05$ kpc may have a bearing on this problem. A
detailed analysis of their characteristics from the Strasbourg
database shows that they are all located in two regions of ongoing
star formation, in Orion and near the star $\rho~Oph$ at distances
$200<r<300$ pc. Both regions are known to exhibit small-scale
variations in extinction coefficient $R_{V}$ and interstellar
extinction. For example, the popular 3D analytical extinction
model by Arenou et al. (1992) for a field in Orion at this
distance gives $A_{V}\approx1^m$, while the 3D model by Gontcharov
(2009, 2012b) gives $A_{V}\approx0.5^m$. Both values are
appreciably higher than the typical $A_{V}$ estimates at these
distances in other directions. It is the large value of the
adopted $A_{V}$ that gives a very blue dereddened color
($(B_{T}-V_{T})_{0}$ for the stars under consideration according
to Eq. (1). As a result, the stars turned out to be well below the
MS in its blue part. It should be noted that the extinction model
and the 3D map of $R_{V}$ variations used give $E(B_{T}-V_{T})$
for these stars that are smaller than other sources of estimates.
Consequently, the derived low luminosity of these stars cannot be
explained by the erroneous extinction and $R_{V}$ estimates and
should be investigated additionally. The most plausible
explanations are: either the peculiarity of the medium and BV
stars on the MS in the star-forming regions causes them to appear
as low-luminosity stars on the H–R diagram, for example, due to an
anomalously high extinction (this explanation appears more
plausible) or they are actually not BV stars but hot sdB subdwarfs
and the medium of the star-forming regions aids the formation of
such stars. Indeed, an especially intense mass loss on the giant
branch, including that due to an external stellar wind at small
distances between the stars, the presence of a close component for
mass transfer, and a high spatial and physical density of clouds
supplying hydrogen into the stellar envelope that has already used
it up contribute to the formation of a hot subdwarf.

The subdwarfs from the OA subsample will be studied in detail
separately after the collection of data on their metallicity.

In the KM subsample in Figs. 7b, 7d, 7f, and 7h, we see the
separation into giants ($\Delta M_{V_{T}}<-2^m$) and dwarfs
($\Delta M_{V_{T}}>-2^m$). The overwhelming majority of both
giants and dwarfs belong to the disk ($e<0.6$, $r_{min}>2$ kpc,
$Z_{max}<2.5$ kpc). Almost all of the remaining ones belong to the
bulge ($r_{min}<2$ kpc). Only two dwarfs have $\Delta
M_{V_{T}}>1^m$ and, thus, are probably cool low-metallicity
subdwarfs in the disk. As expected, the KM subsample contains
almost no cool subdwarfs due to selection. However, the tendency
for $\Delta M_{V_{T}}$ to increase with increasing $e$ and
decreasing $r_{min}$ caused by a drop in mean metallicity should
be noted for both bulge giants and dwarfs ($e>0.6$, $r_{min}<2$
kpc). It can be seen that this tendency is opposite to that for
the OA subsample and reflects a different nature of the
low-luminosity stars: these are dwarfs and branch giants in the KM
subsample and giants of the blue part of the horizontal branch in
the OA subsample.

In Fig. 8, which is similar to Fig. 7, we will analyze the entire
FG subsample (see Figs. 8a, 8c, 8e, 8g) and the GCS stars from
this subsample (see Figs. 8b, 8d, 8f, 8h). It can be seen that
with regard to the disk stars ($e<0.3$, $r_{min}>3$ kpc,
$Z_{max}<2$ kpc), the distribution of the subsample stars in the
figure virtually coincides with that of the GCS stars. However,
with regard to the remaining stars, the subsample is more complete
than theGCS.Dwarfs ($\Delta M_{V_{T}}>-2^m$) and giants ($\Delta
M_{V_{T}}<-2^m$) with their perigalacticons in the bulge
($r_{min}<2$ kpc) and apogalacticons in the halo ($Z_{max}>2$ kpc)
dominate among the stars outside the disk. In contrast to the disk
stars with $\overline{\Delta M_{V_{T}}}<0^m$, they have a low
luminosity ($\overline{\Delta M_{V_{T}}}>0^m$). This makes them
similar to the KM subsample stars and distinguishes them from the
hot subdwarfs of the OA subsample.

There are five halo stars with their perigalacticons outside the
bulge in the FG subsample: the previously mentioned halo group.
Their small number in the subsample is explained by the fact that
the FG dwarfs have a low luminosity and are not visible from afar,
while the horizontal-branch giants with
($(B_{T}-V_{T})_{0}<0.85^m$, i.e., with $Fe/H<-0.6$, are very few.
However, halo giants are absent altogether in the KM subsample,
because only giants with $Fe/H>-0.6$ may turn out to be in the
interval $(B_{T}-V_{T})_{0}>0.85^m$, and there are probably no
such stars in the halo. Out of these five stars, HIP 103311 is a
dwarf, while HIP 92 167, 85 855, 71 458, and 62 747 are
horizontal-branch low-metallicity giants. All of their
characteristics confirm that they belong to the halo, with the
interval $-5.2^m<\Delta M_{V_{T}}<-2.9^m$ populated by these stars
just corresponding to the location of the horizontal branch
relative to the MS for the F and G types. Therefore, the entire
isolated group of stars with $-5.2^m<\Delta M_{V_{T}}<-2.4^m$,
$Z_{max}>2$ kpc, and $r_{max}>12$ kpc probably consists of
horizontal-branch low-metallicity giants. It can be seen from Fig.
8 that these stars are absent in the GCS and were first detected
as an isolated group in this study.

Gontcharov et al. (2011) found no cool disk subdwarf (with
$e<0.4$). The GCS also gives few stars with $e<0.4$ and $\Delta
M_{V_{T}}>1$. The FG subsample contains several such stars.
Analysis of their characteristics from the Strasbourg database
shows that they are all young nonsingle and variable stars in
starforming regions rather than subdwarfs.

\section{CONCLUSIONS}

We analyzed the Galactic orbits of 27 440 stars of all classes
with accurate $\alpha$, $\delta$ and $\pi>3$ mas from the
Hipparcos catalogue,  $\mu$ from the Tycho-2 catalogue, and
$V_{r}$ from the Pulkovo Compilation of Radial Velocities (PCRV).
The detection of systematic errors in $V_{r}$ when the PCRV was
created and the noticeable difference between . from Tycho-2 and
Hipparcos caused by the orbital motions in star pairs forced us to
estimate the influence of these errors on orbital parameters: the
peri- and apogalactic distances $r_{min}$ and $r_{max}$ , the
eccentricity $e$, and the largest distance of the orbit from the
Galactic plane $Z_{max}$. We found that the errors of $\mu$ due to
the duplicity of stars are tangible only in the statistics of
orbital parameters for very small samples (fewer than 10 stars),
while the errors of the radial velocities are noticeable in the
statistics of orbital parameters for stars far from the Sun, i.e.,
halo stars. The sample considered here is much more representative
than the Geneva–Copenhagen survey with regard to F–G stars that do
not belong to the disk and exceeds considerably any other sample
for O–A and K–M stars. This allows our study to be considered as
the largest survey of Galactic orbits in the solar neighborhood to
date. Note that the derived orbital parameters agree with those
from the GCS for the same stars.

Here, we analyzed the distribution of stars in the
multidimensional space of orbital parameters, dereddened colors,
and absolute magnitudes so far almost without invoking the stellar
metallicities and ages (this will be done in a subsequent paper).
Since the sample is limited in parallax and apparent magnitude,
many of the groups of stars (for example, halo stars with circular
orbits) cannot appear in it. However, even our analysis of
selection-free regions of this multidimensional space allowed us
to establish a nonuniformity of the distribution of stars in it
and to identify several groups. First of all, this study allowed
the radius of the bulge to be determined (2 kpc) and showed that
the bulge and the halo are not homogeneous subsystems of the
Galaxy. The stars with their perigalacticons in the bulge are
clearly divided into stars with their apogalacticons in the halo
and the disk, while the stars with their apogalacticons in the
halo are divided into stars with their perigalacticons in the
bulge and the disk. Thus, instead of the evidence for the
membership of a star in a subsystem, the evidence for which
subsystem the perigalacticon and apogalacticon of its orbit are
located in is more informative. Therein may lie the difference in
the origin of stars: in the Galaxy or in the accreted satellites.

The nonuniformity of the distribution of bulge stars with their
apogalacticons in the disk in apogalactic distances may be
explained by the dynamical association of these stars with spiral
arms.

Our investigation showed that using Galactic orbits is promising.
In future, they can be applied to calculate the statistical
characteristics of the disk and, by invoking the stellar
metallicities and ages, for a detailed analysis of the bulge and
the halo.

\section{ACKNOWLEDGMENTS}

We used resources from the Strasbourg Astronomical Data Center
(Centre de Donn\'ees astronomiques de Strasbourg). This study was
supported by Program P21 of the Presidium of the Russian Academy
of Sciences and the Ministry of Education and Science of the
Russian Federation under contract no. 8417.

\section{REFERENCES}

\hskip 0.625 cm 1. F. Arenou,M. Grenon, and A. Gomez, Astron.
Astrophys. 258, 104 (1992).

2. T. Bensby, J. C. Yee, S. Feltzing, et al., Astron. Astrophys.

3. V. Bobylev and A. Bajkova, Mon. Not. R. Astron. Soc. 408, 1788
(2010).

4. J. Bovy, H.-W. Rix, and D. W. Hogg, Astrophys. J. 751, 131
(2012).

5. A. Bressan, P. Marigo, L. Girardi, et al., Mon. Not. R. Astron.
Soc. 427, 127 (2012).

6. B. T. Draine, Ann. Rev. Astron. Astrophys. 41, 241 (2003).

7. B. Famaey, A. Jorissen, X. Luri, et al., Astron. Astrophys.
430, 165 (2005).

8. M. Fellhauer, V. Belokurov, N. W. Evans, et al., Astrophys. J.
651, 167 (2006).

9. G. A.Gontcharov andO. V. Kiyaeva, Astron. Lett. 28, 261 (2002).

10. G. A. Gontcharov, Astron. Lett. 32, 759 (2006).

11. G. A. Gontcharov, Astron. Lett. 35, 780 (2009).

12. G. A. Gontcharov, Astron. Lett. 38, 12 (2012a).

13. G. A. Gontcharov, Astron. Lett. 38, 87 (2012b).

14. G. A. Gontcharov, Astron. Lett. 38, 694 (2012c).

15. G. A. Gontcharov, Astron. Lett. 38, 771 (2012d).

16. G. A. Gontcharov, A. A. Andronova, O. A. Titov, et al.,
Astron. Astrophys. 365, 222 (2001).

17. G.A.Gontcharov,A. T. Bajkova, P. N. Fedorov, et al., Mon. Not.
R. Astron. Soc. 413, 1581 (2011).

18. A. Helmi, J. F. Navarro, B. Nordstr . om, et al., Mon. Not. R.
Astron. Soc. 365, 1309 (2006).

19. L. Hernquist, Astrophys. J. 356, 359 (1990).

20. E. Hog, C. Fabricius, V. V. Makarov, et al., Astron.
Astrophys. 355, L27 (2000).

21. J. Holmberg, B. Nordstr . om, and J. Andersen, Astron.
Astrophys. 475, 519 (2007).

22. J. Holmberg, B. Nordstr . om, and J. Andersen, Astron.
Astrophys. 501, 941 (2009).

23. F. van Leeuwen, Astron. Astrophys. 474, 653 (2007).

24. J. R. de Medeiros and M. Mayor, Astron. Astrophys. Suppl. Ser.
139, 433 (1999).

25. M. Miyamoto and R. Nagai, Publ. Astron. Soc. Jpn. 27, 533
(1975).

26. B. Nordstr . om, M. Mayor, J. Andersen, et al., Astron.
Astrophys. 418, 989 (2004).

27. M. Perryman, Astronomical Applications of Astrometry
(Cambridge Univ. Press, Cambridge, 2009).

28. R. Sch . onrich, J. Binney, and W. Dehnen, Mon. Not. R.
Astron. Soc. 403, 1829 (2010).

29. V. V. Vityazev, V. V. Bobylev, and G. A. Gontcharov, Vestn.
SPb. Univ., Ser. 1, No. 4 (25), 111 (2003). 549, 147 (2013).

\begin{figure}
\includegraphics{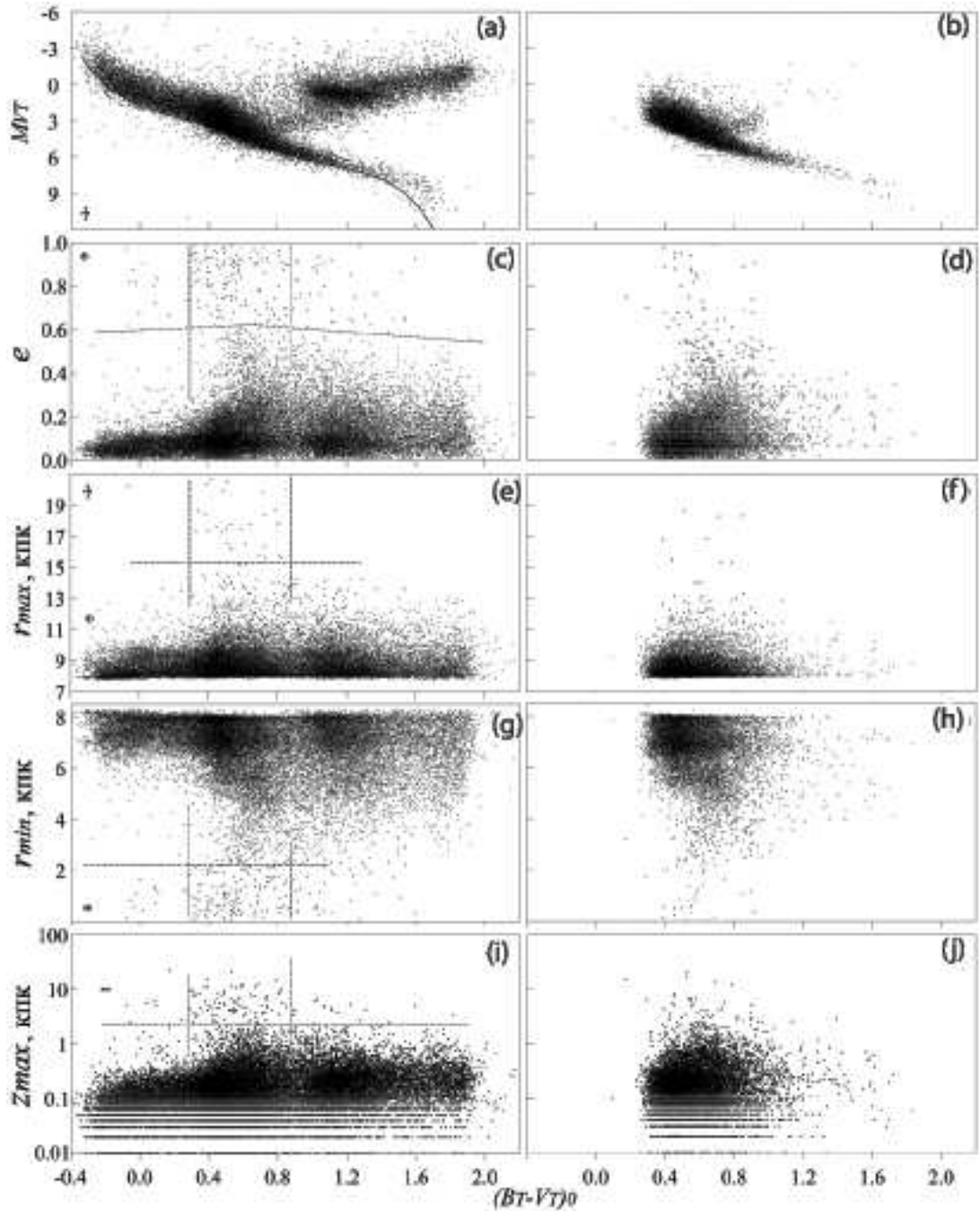}
\caption{``$(B_{T}-V_{T})_{0}$ -- $M_{V_{T}}$'' diagram for the
sample (a) and GCS (b) stars (the line indicates the theoretical
ZAMS); ``$(B_{T}-V_{T})_{0}$ -- $e$'' for the sample (c) and GCS
(d) stars; ``$(B_{T}-V_{T})_{0}$ -- $r_{max}$'' for the sample (e)
and GCS (f) stars; ``$(B_{T}-V_{T})_{0}$ -- $r_{min}$'' for the
sample (g) and GCS (h) stars; ``$(B_{T}-V_{T})_{0}$ -- $Z_{max}$''
for the sample (i) and GCS (j) stars. The crosses near the
vertical axes indicate typical errors in the data for a star. The
dashed lines are discussed in the text. } \label{hr}
\end{figure}

\begin{figure}
\includegraphics{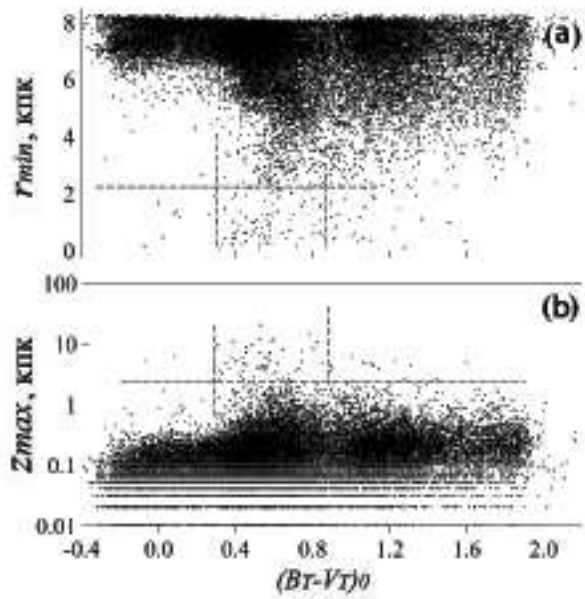}
\caption{``$(B_{T}-V_{T})_{0}$ -- $r_{min}$'' (a) and
``$(B_{T}-V_{T})_{0}$ -- $Z_{max}$'' (b) diagrams for 25 082
sample stars with $\sigma(\pi)/\pi<0.2$,
$\sigma(\mu_{\alpha}\cos\delta)<3$, and $\sigma(\mu_{\delta})<3$
mas yr$^{-1}$. The dashed lines were taken from Fig.~1.}
\label{pi02}
\end{figure}

\begin{figure}
\includegraphics{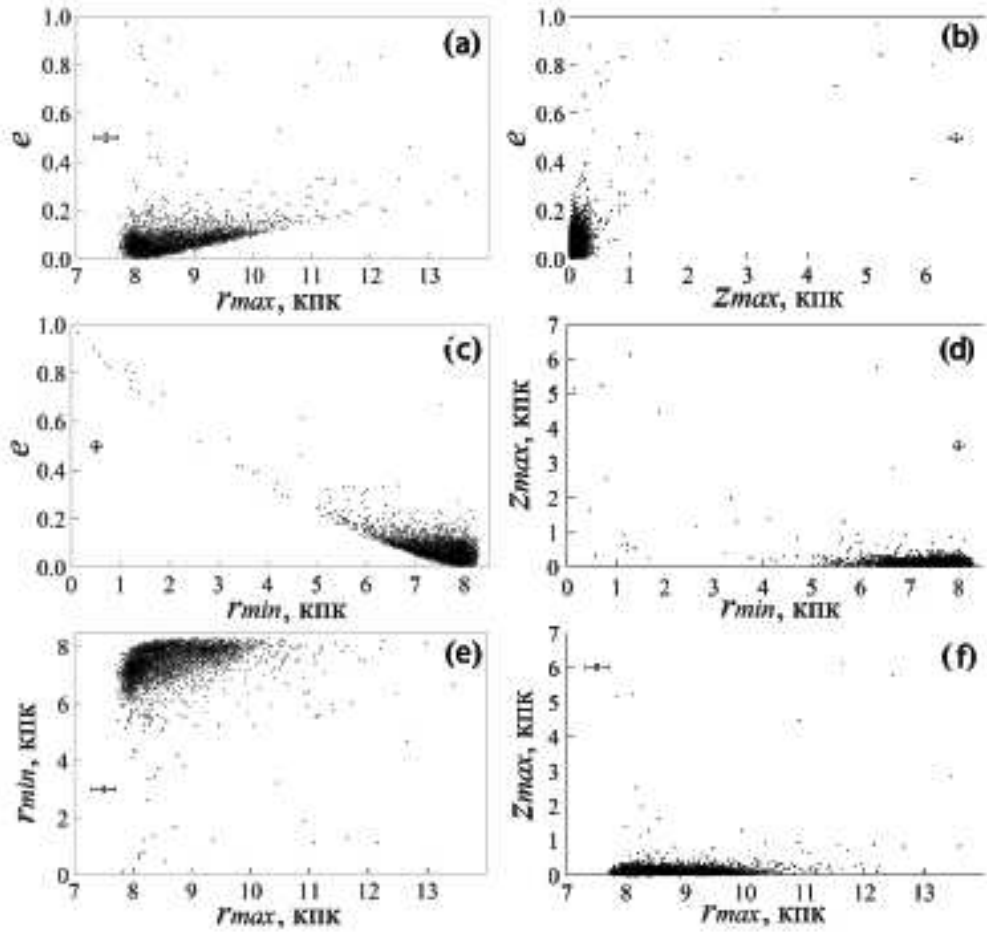}
\caption{Positions of 5377 stars from the OA subsample on the (a)
``$r_{max}$ -- $e$'', (b) ``$Z_{max}$ -- $e$'', (c) ``$r_{min}$ --
$e$'', (d) ``$r_{min}$ -- $Z_{max}$'', (e) ``$r_{max}$ --
$r_{min}$'', and (f) ``$r_{max}$ -- $Z_{max}$'' diagrams. The
crosses indicate typical errors for a star. } \label{oa}
\end{figure}

\begin{figure}
\includegraphics{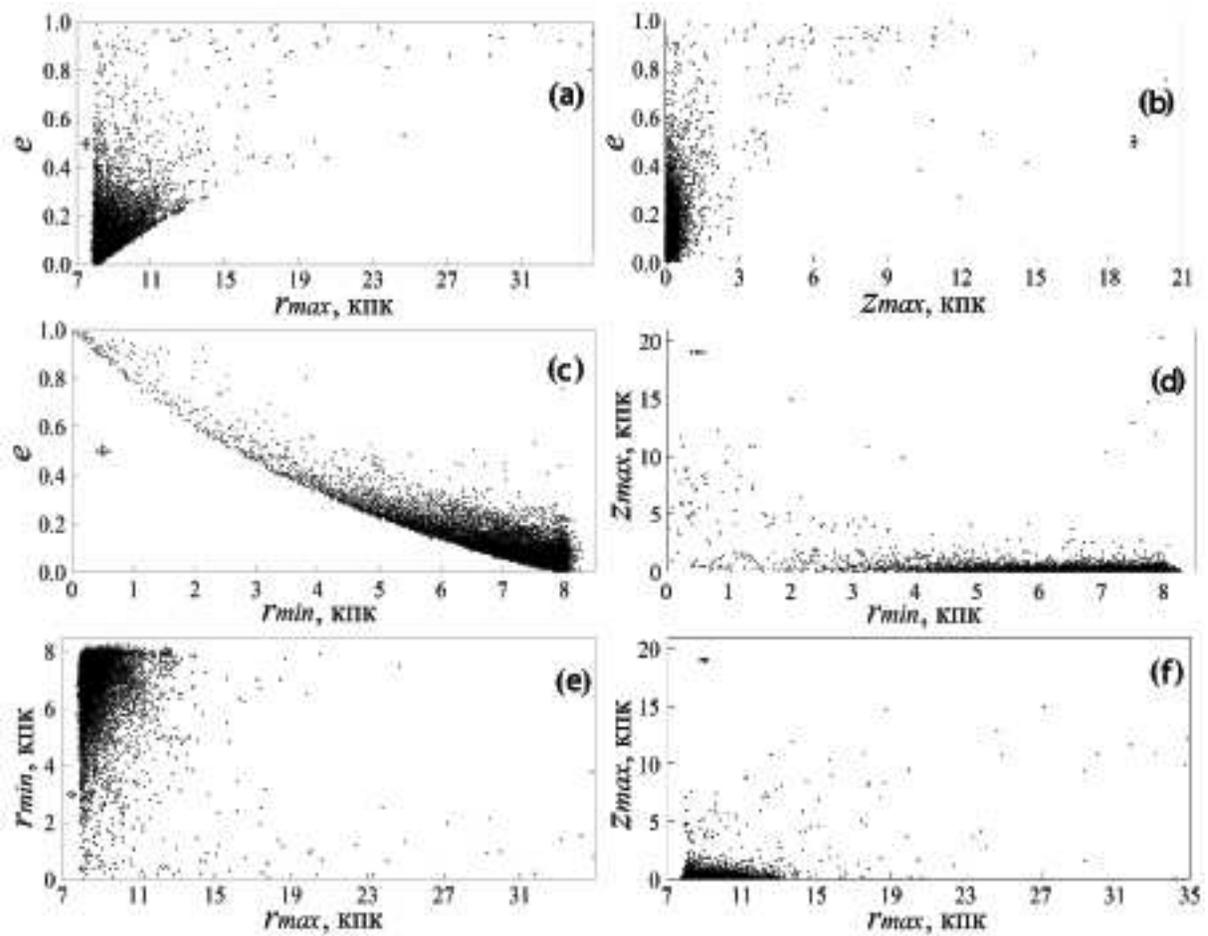}
\caption{Same as Fig. 3 for 12 600 stars from the FG subsample. }
\label{fg}
\end{figure}

\begin{figure}
\includegraphics{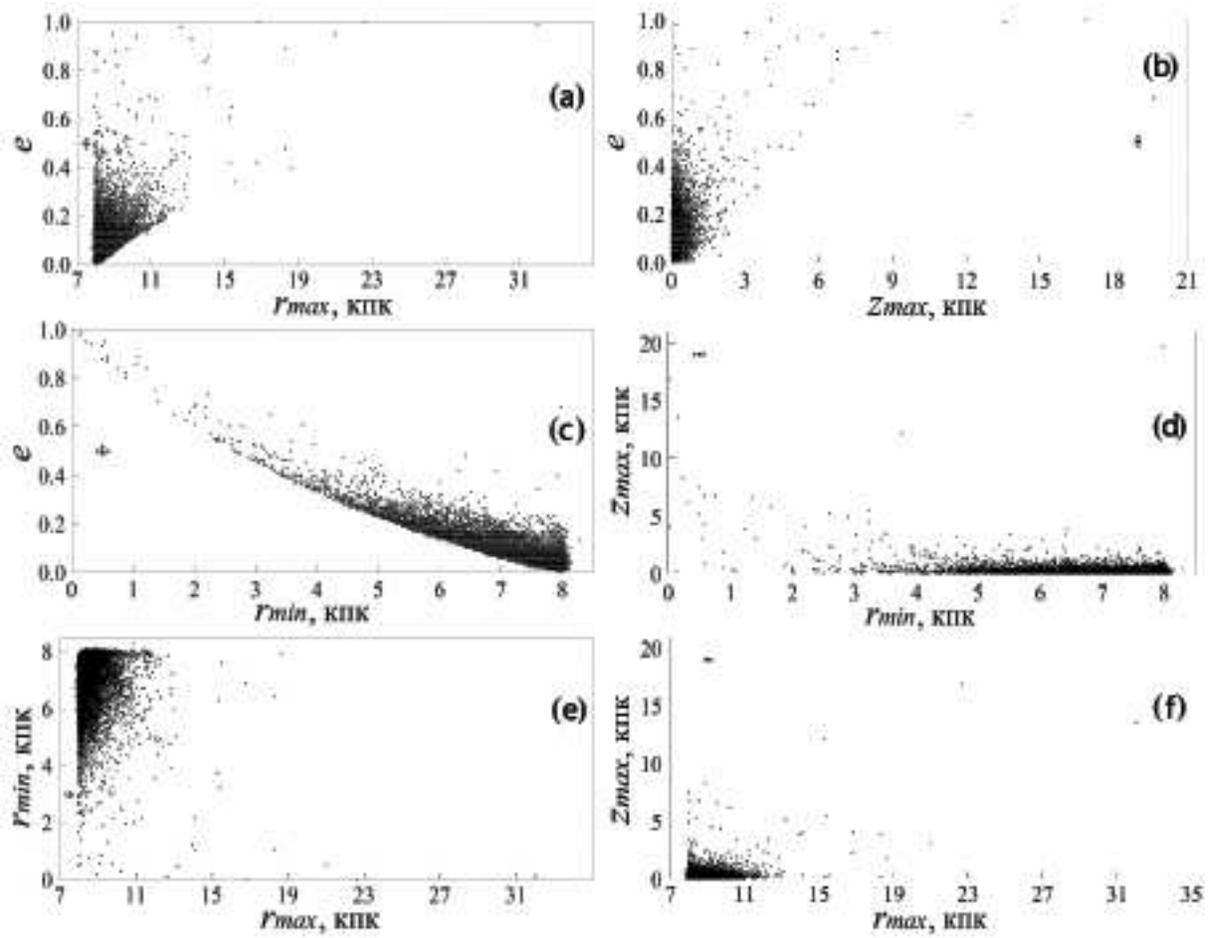}
\caption{Same as Fig. 3 for the GCS stars from the FG subsample.}
\label{fggcs}
\end{figure}

\begin{figure}
\includegraphics{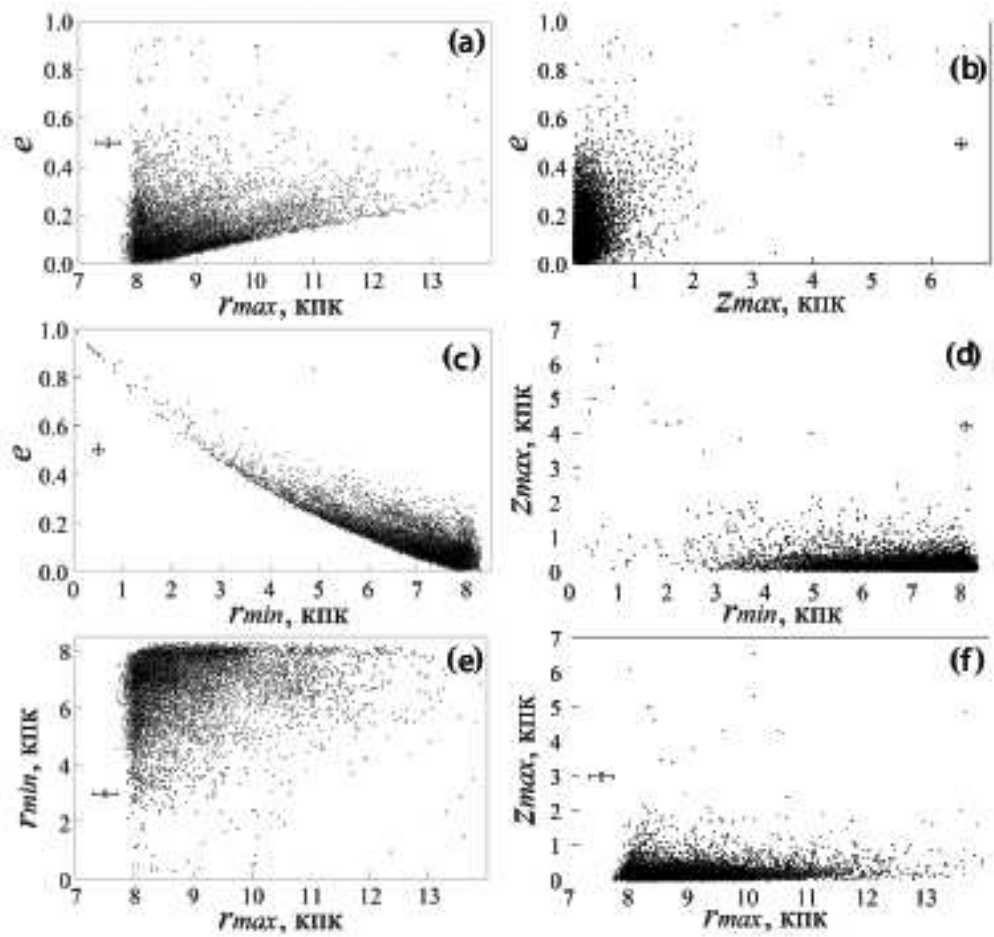}
\caption{Same as Fig. 3 for 9463 stars from the KM subsample. }
\label{km}
\end{figure}

\begin{figure}
\includegraphics{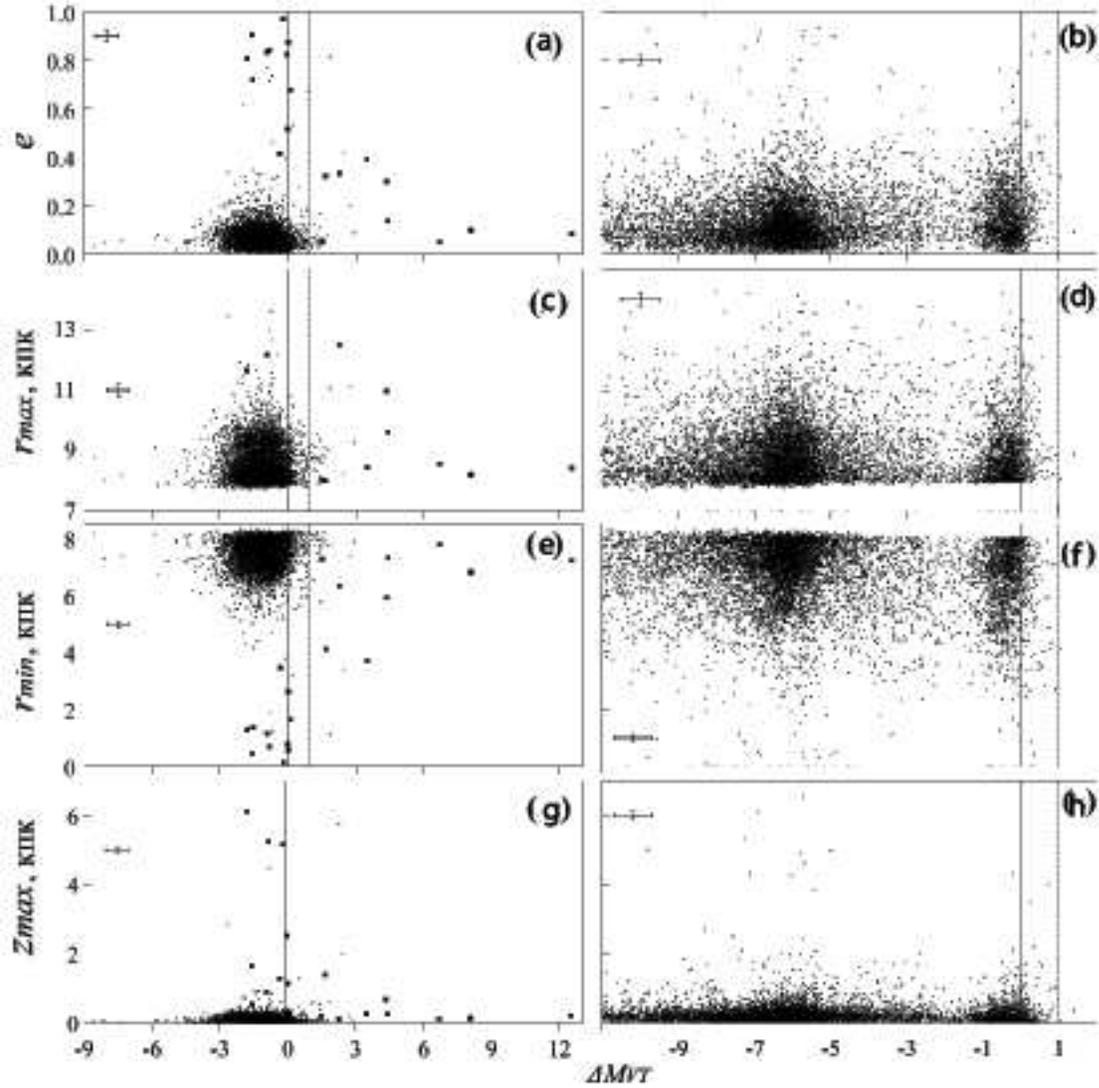}
\caption{Positions of 5377 stars from the OA subsample and 9463
stars from the KM subsample on the ``$\Delta M_{V_{T}}$ -- $e$''
(a) and (b), `$\Delta M_{V_{T}}$ -- $r_{max}$'' (c) and (d),
`$\Delta M_{V_{T}}$ -- $r_{min}$'' (e) and (f), ``$\Delta
M_{V_{T}}$ -- $Z_{max}$'' (g) and (h) diagrams. The solid vertical
straight line marks $\Delta M_{V_{T}}=0^m$ and the dashed straight
line marks $\Delta M_{V_{T}}=1^M$, which corresponds to a
$2\sigma$ error in calculating $M_{V_{T}}$. The crosses indicate
typical errors for a star. } \label{offset1}
\end{figure}

\begin{figure}
\includegraphics{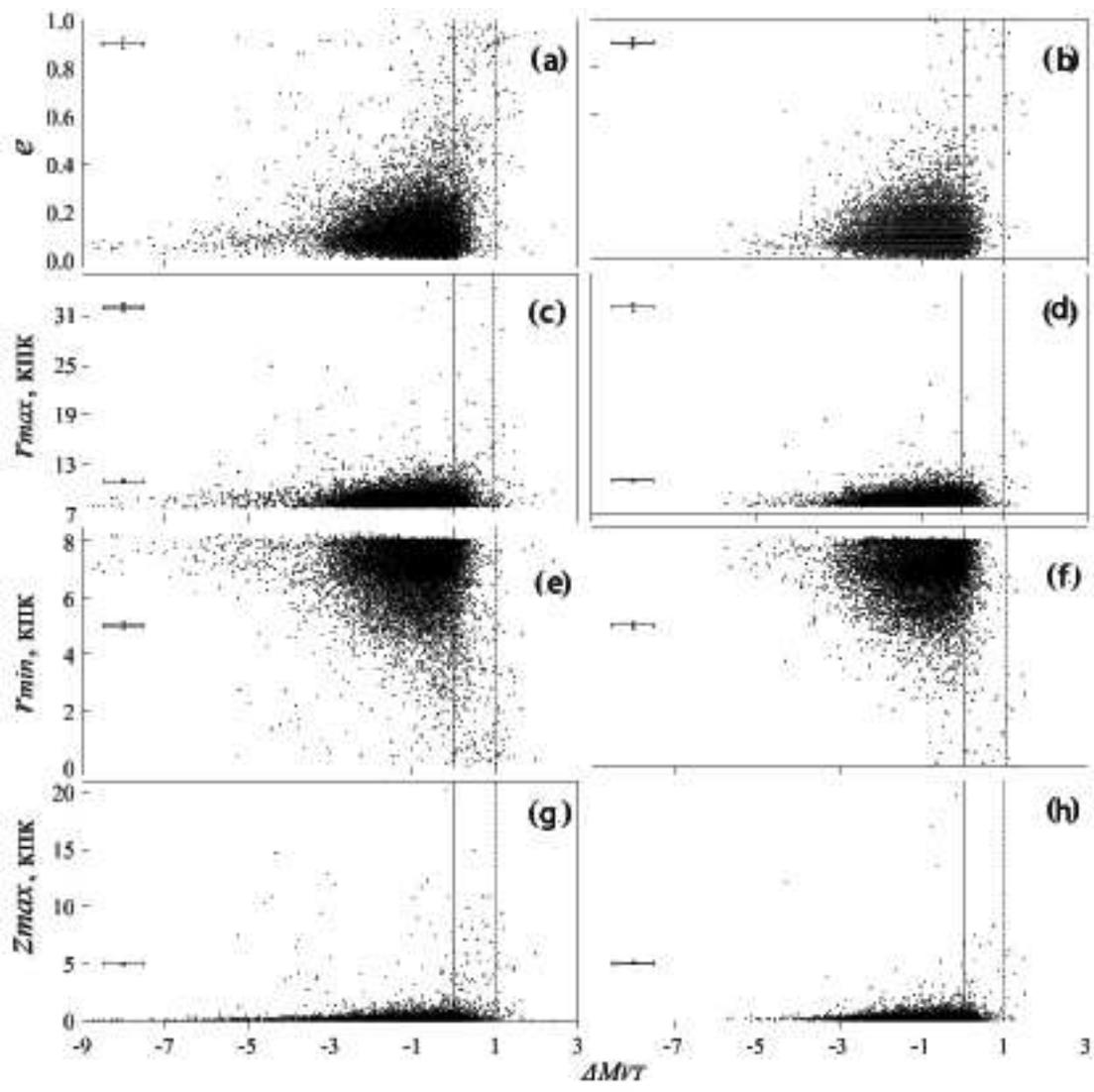}
\caption{Same as Fig. 7 for 12 600 stars from the FG subsample and
GCS stars from the FG subsample. } \label{offset2}
\end{figure}

\end{document}